\documentclass[aps,prb,twocolumn,groupedaddress,citeautoscript]{revtex4}

\usepackage{graphicx}
\usepackage{amsmath}
\usepackage{times}
\usepackage{colordvi}

\graphicspath{{.}{./EPS/}}

\begin{document}

\title{Spin interferometry with electrons in nanostructures: A
road to spintronic devices}

\author{U. Z\"ulicke}
\affiliation{Institute of Fundamental Sciences, Massey University,
Private Bag 11~222, Palmerston North, New Zealand}

\date{version of 17 July 2004}

\begin{abstract}

The wave nature of electrons in semiconductor nanostructures
results in spatial interference effects similar to those exhibited
by coherent light. The presence of spin--orbit coupling renders
interference in spin space and in real space interdependent,
making it possible to manipulate the electron's spin state by
addressing its orbital degree of freedom. This suggests the
utility of electronic analogs of optical interferometers as
blueprints for new spintronics devices. We demonstrate the
usefulness of this concept using the Mach--Zehnder interferometer
as an example. Its spin--dependent analog realizes a
spin--controlled field--effect transistor without magnetic
contacts and may be used as a quantum logical gate.

\end{abstract}


\maketitle

Quantum phase coherence of electrons in nanostructures has been
exhibited in a number of interference experiments. Aharonov--Bohm
oscillations of the electrical conductance through mesoscopic
rings were observed~\cite{aboscmetal,aboscsemi} and used to design
the first solid-state electron interferometers~\cite
{bartinterfere,tuneabosc}. Electronic double--slit~\cite
{amir:prl:94} and Mach--Zehnder~\cite{electronMZ} interferometers
have recently been realized in two--dimensional (2D) semiconductor
heterostructures~\cite{hetero1}. In addition to revealing
intriguing properties of matter at the nanoscale,
quantum--coherence effects could possibly be utilized for the
creation of new electronic devices. One example are
proposals~\cite{dircoupl1,dircoupl2} to build quantum switches
from coupled electron wave guides.

Recent ideas~\cite{sciencerev,lossbook} to manipulate current flow
by addressing the spin degree of freedom of charge carriers 
are attracting a lot of interest. Some of these proposals
involve phase--coherent spin--dependent transport in
nanostructures. 
Spin dependence can be
introduced by the presence of magnetic materials, as in the
spin--dependent Fabry--P\'erot
interferometer\cite{spininterfere1,spininterfere2}
realized with electrons in magnetic multilayers.
Similarly, an early theoretical suggestion~\cite{spinfet} for an
electronic analog of the electro--optical modulator utilized
magnetic contacts as polarizers and analyzers. However, its basic
functionality rests on the experimentally confirmed~\cite
{nitta:prl:97,schaep:prb-rc:97,syoji:jap:01} tunability of spin
precession due to the Rashba effect~\cite{byra:jpc:84} that arises
from structural inversion asymmetry~\cite{wink:prb:00} in 2D
electron systems. As a possible alternative to an entirely
magnet--based spintronics, theoretical proposals for spin
polarizers without magnets have been put forward recently which
are based on the interplay of the Rashba effect with resonant
tunneling~\cite{andra:prb:99,kis:apl:01,koga:prl:02a,
uz:prb-rc:02a} and quantum interference~\cite{nitta:apl:99,
irene:prb-rc:03,kis:jap:03}. Here
we explore a new direction toward spintronics devices that are
inspired by quantum--optics setups and combine interferometry in
real space with spin precession in nonmagnetic nanostructures.

\begin{figure}[b]
\centerline{\includegraphics[width=2.2in]{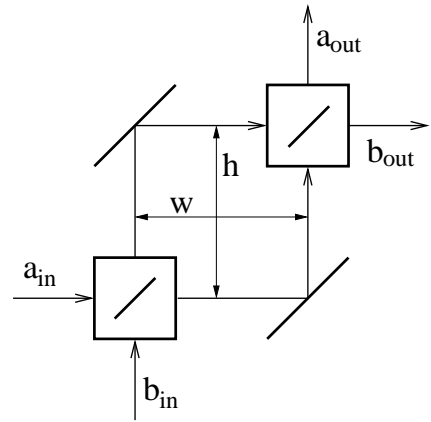}}
\caption{Schematic setup of the spin--dependent electronic
Mach--Zehnder interferometer. Two incident electron spinors,
denoted as $a_{\text{in}}=(a_{\text{in}+},a_{\text{in}-})^{\text
{T}}$ and $b_{\text{in}}=(b_{\text{in}+}, b_{\text{in}-})^{\text
{T}}$, are mixed by a beam splitter and fed into two
interferometer arms. Externally tunable Rashba spin splitting
induces spin--dependent dynamical phase shifts for traveling
electron waves. The interplay between spin precession during
propagation and interference at the second beam splitter
determines the output spinor amplitudes $a_{\text{out}}$,
$b_{\text{out}}$.
\label{mzsetup}}
\end{figure}
As an example, we consider the geometry of an electronic
Mach--Zehnder (MZ) interferometer shown in Fig.~\ref{mzsetup}. The
setup is analogous to its quantum--optics counterpart~\cite
{loudon}. Two electronic beam splitters and perfectly reflecting
`mirrors' can be realized in a suitably nanostructured 2D electron
systems~\cite{liu:sci:99,basel:sci:99,electronMZ}, e.g., by point
contacts and a hard--wall confinement. To be specific, we assume
the interferometer arms to be quasi--onedimensional electron wave
guides and characterize beam splitters and mirrors by appropriate
scattering matrices~\cite{datta}. Spin--dependent quantum input
and output amplitudes are labeled according to
Fig.~\ref{mzsetup}. In addition, front and back gates are used to
independently manipulate Rashba spin splitting and electron
density in the 2D electron system~\cite{shay:sci:99}. In the
following, we neglect electron--electron interactions and assume
low enough temperature such that the phase--coherence length
exceeds the interferometer size.

\begin{figure}[t]
\centerline{\includegraphics[width=2in]{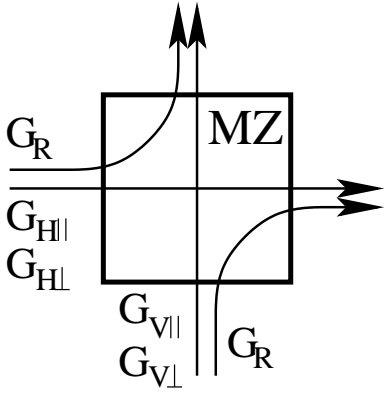}}
\caption{Transport coefficients for the spin--dependent MZ
interferometer, given in the basis of spin--split
eigenstates in each lead. Reflection processes are described by a
spin--independent conductance $G_{\text{R}}$. In contrast,
horizontal transmission depends on whether spin is conserved
($G_{\text{H}\parallel}$) or flipped ($G_{\text{H}\perp}$). The
same holds for vertical transmission. All conductances can
be modulated by gate voltages.
\label{conductmat}}
\end{figure}
We first illustrate the basic function of the spin--dependent MZ
interferometer by considering the ideal case where the width of
electron wave guides is much smaller than the spin--orbit--induced
spin precession length $L_{\text{so}}$, and only the lowest
wave--guide subband is occupied. To a good approximation,
single--electron eigenstates are then also eigenstates of the
spin component that is perpendicular to the wave guide and lies in
the plane of the 2D electron system. For a given energy $E$,
eigenstates labeled by $\sigma=\pm$ have wave numbers
\begin{equation}
k_\sigma = k_E - \sigma\, \frac{\pi}{L_{\text{so}}}\,\mbox{ where
 } \, k_E = \sqrt{\frac{2 m E}{\hbar^2}+\left(\frac{\pi}{L_{\text
{so}}} \right)^2} \,  .
\end{equation}
The requirement of mutually perpendicular propagation direction
and spin--quantization axis for eigenstates results in unavoidable
spin mixing at beam splitters and mirrors. In addition, the
two eigenstates acquire different dynamical phases during
propagation in the same interferometer arm. Interference effects
in the structure are most conveniently embodied in an effective
spin--resolved scattering matrix
\cite{pareek}
${\mathcal S}$, relating output spinors $a_{\text{out}}$,
$b_{\text{out}}$ to the incident ones $a_{\text{in}}$,
$b_{\text{in}}$:
\begin{equation}
\left(\begin{array}{c} a_{\text{out}+} \\ a_{\text{out}-} \\
b_{\text{out}+} \\ b_{\text{out}-} \end{array} \right) =
\underbrace{\left(\begin{array}{cccc} r_{1++} & r_{1+-} & t_{2
++} & t_{2+-} \\ r_{1-+} & r_{1--} & t_{2-+} & t_{2--} \\ t_{1
++} & t_{1+-} & r_{2++} & r_{2+-} \\ t_{1-+} & t_{1--} & r_{2-+} &
r_{2--} \end{array} \right)}_{\mathcal S}
\left(\begin{array}{c} a_{\text{in}+} \\ a_{\text{in}-} \\
b_{\text{in}+} \\ b_{\text{in}-} \end{array} \right) .
\end{equation}
It can be calculated from the individual scattering matrices for
the beam splitters (${\mathcal S}_{\text{bs}}$) and mirrors
(${\mathcal S}_{\text{m}}$). These are~\cite{rashbascatt}
\begin{equation}
{\mathcal S}_{\text{bs}}=\left(\begin{array}{cccc} \frac{i}{2} &
-\frac{1}{2} & \frac{1}{\sqrt{2}} & 0 \\ -\frac{1}{2} & \frac{i}
{2} & 0 & \frac{1}{\sqrt{2}} \\ \frac{1}{\sqrt{2}} & 0 &\frac{i}
{2} & \frac{1}{2} \\ 0 & \frac{1}{\sqrt{2}} & \frac{1}{2} & \frac
{i}{2} \end{array}\right); \,
{\mathcal S}_{\text{m}}=\frac{1}{\sqrt{2}} \left(\begin{array}
{cccc} i & -1 & 0 & 0 \\ -1 & i & 0 & 0 \\ 0 & 0 & i & 1 \\ 0 & 0
& 1  & i \end{array}\right) ,
\end{equation}
where the optimal case of identical and symmetric beam splitters
is assumed. A standard calculation yields ${\mathcal S}= e^{i k_E
(h+w)} {\mathcal S}^{\text{(MZ)}}$ with the energy dependence
entirely contained in $k_E$. ${\mathcal S}^{\text{(MZ)}}$ has
reflection and transmission amplitudes
\begin{subequations}\label{coefficients}
\begin{eqnarray}
r^{\text{(MZ)}}_{j\sigma\sigma^\prime} &=& i r^{\text{(MZ)}}_{1
\sigma,-\sigma}=-i r^{\text{(MZ)}}_{2\sigma,-\sigma}=\frac{\sin
\left(\frac{\pi h}{L_{\text{so}}}\right)\sin\left(\frac{\pi w}
{L_{\text{so}}}\right)}{i \sqrt{2}} , \\
t^{\text{(MZ)}}_{2++}&=&\left[t^{\text{(MZ)}}_{2--}\right]^\ast =
- e^{-i\frac{\pi h}{L_{\text{so}}}}\cos\left(\frac{\pi w}{L_{\text
{so}}}\right) \quad , \\
t^{\text{(MZ)}}_{1++}&=&\left[t^{\text{(MZ)}}_{1--}\right]^\ast =
- e^{-i\frac{\pi w}{L_{\text{so}}}}\cos\left(\frac{\pi h}{L_{\text
{so}}}\right) \quad ,\\
t^{\text{(MZ)}}_{2+-}&=&-t^{\text{(MZ)}}_{2-+} = \sin\left(\frac
{\pi w}{L_{\text{so}}}\right)\cos\left(\frac{\pi h}{L_{\text{so}}}
\right) \quad , \\
t^{\text{(MZ)}}_{1+-}&=&-t^{\text{(MZ)}}_{1-+} = - \sin\left(\frac
{\pi h}{L_{\text{so}}}\right)\cos\left(\frac{\pi w}{L_{\text{so}}}
\right) \quad .
\end{eqnarray}
\end{subequations}

An important functional aspect of the spin MZ interferometer can
be gleaned from the linear $4\times 4$ conductance matrix
$\mathcal G$ that is related to the scattering matrix via
${\mathcal G}_{j k}=\frac{e^2}{2\pi\hbar} |{\mathcal S}_{j k}|^2$.
It has a surprisingly simple form:
\begin{equation}
{\mathcal G}=\left(\begin{array}{cccc} G_{\text{R}} & G_{\text{R}}
& G_{\text{V}\parallel} & G_{\text{V}\perp} \\ G_{\text{R}} &
G_{\text{R}} & G_{\text{V}\perp} & G_{\text{V}\parallel} \\
G_{\text{H}\parallel} & G_{\text{H}\perp} & G_{\text{R}} &
G_{\text{R}} \\ G_{\text{H}\perp} & G_{\text{H}\parallel} &
G_{\text{R}} & G_{\text{R}} \end{array}\right) \quad ,
\end{equation}
and we find from the results given in Eqs.~(\ref{coefficients})
\begin{subequations}
\begin{eqnarray}
G_{\text{R}} &=& \frac{1}{8}\left[ 1 - \cos\left(\frac{2 \pi w}
{L_{\text{so}}}\right) \right]\left[ 1 - \cos\left(\frac{2 \pi h}
{L_{\text{so}}}\right)\right] \, , \\
G_{\text{V}\parallel} &=& \frac{1}{2} \left[1 + \cos\left(\frac{2
\pi w}{L_{\text{so}}}\right) \right] \quad , \\
G_{\text{H}\parallel} &=& \frac{1}{2} \left[1 + \cos\left(\frac{2
\pi h}{L_{\text{so}}}\right) \right] \quad , \\
G_{\text{V}\perp} &=& \frac{1}{4}\left[ 1 - \cos\left(\frac{2 \pi
w}{L_{\text{so}}}\right) \right]\left[ 1 + \cos\left(\frac{2 \pi
h}{L_{\text{so}}}\right)\right]\, , \\
G_{\text{H}\perp} &=& \frac{1}{4}\left[ 1 + \cos\left(\frac{2 \pi
w}{L_{\text{so}}}\right) \right]\left[ 1 - \cos\left(\frac{2 \pi
h}{L_{\text{so}}}\right)\right]\, .
\end{eqnarray}
\end{subequations}
The physical meaning of these conductances can be understood in
terms of probabilities for associated reflection and transmission
processes, as illustrated in Fig.~\ref{conductmat}. Interestingly,
reflection processes turn out to be characterized by a
spin--rotationally invariant effective conductance $G_{\text{R}}$.
In contrast, transmission probabilities depend on whether or not
the spin state is conserved. Spin--conserving transmission depends
only on the interferometer size in propagation direction and is
different, in general, from transmission involving a spin flip.
All conductance coefficients oscillate as functions of the
spin--precession length. In addition to general sum rules that are
mandated by current conservation~\cite{datta}, they obey the
relation
\begin{equation}
2 G_{\text{R}} = \frac{G_{\text{V}\perp}G_{\text{H}\perp}}
{G_{\text{V}\parallel}G_{\text{H}\parallel}} \quad .
\end{equation}
Within the single--subband approximation considered here,
conductance coefficients are independent of electron energy. This
is the same property that the ideal spin field--effect transistor
suggested by \citet{spinfet} exhibits. Hence all electrons that
are injected into the spin--dependent MZ interferometer within a
particular energy window opened by a finite bias voltage will be
scattered in an identical manner.

While knowledge of the general conductance coefficients given
above allows us to predict transport for interferometers of any
size, it is useful to highlight a few special cases. (i)~When both
$w$ and $h$ are integer multiples of the spin--precession length
$L_{\text{so}}$, transmission through the MZ interferometer is
perfect, and the vertical and horizontal channels are completely
decoupled. Spin--split eigenstates acquire phase factors depending
on their wave--number difference and the interferometer size in
propagation direction. In other words, the interferometer acts as
if it were not there. (ii)~When $h$ is an integer multiple of
$L_{\text{so}}$ and $w$ a half--integer multiple, transmission is
still perfect but incurs a spin flip in the vertical channel. This
is true independent of the actual spin state of electrons that are
injected into the interferometer. For symmetry reasons, an
analogous reflectionless case exists for $h$ and $w$ being
half--integer and integer multiples of $L_{\text{so}}$,
respectively; but then the spin flip occurs in the horizontal
channel. (iii)~A purely reflecting case is realized when both $h$
and $w$ are half--integer multiples of $L_{\text{so}}$. The spin
of incoming electrons turns out to be conserved in this reflection
process. Hence electrons incident in an eigenstate cannot be in an
eigenstate after reflection, and their spin will start precessing.

Clearly a suitably designed structure would enable switching
between any two cases discussed above by adjusting the
spin--precession length. For example, in a MZ interferometer with
$h=w$, there would be two voltages realizing cases (i) and (iii),
respectively. As function of one input channel only, such a device
acts as a voltage--controlled switch or spin field--effect
transistor, similar to the one discussed in
Ref.~\onlinecite{spinfet}. As opposed to the design by Datta and
Das, however, no magnetic contacts are involved in the present
setup, which is somewhat related to the spin interference device
proposed in Ref.~\onlinecite{nitta:apl:99}. In fact, while
spin--dependent interference is crucial for the possibility to
switch between perfect transmission and reflection, its effect on
the incoming electron beam is the same for any polarization, in
particular also for an unpolarized beam.

Changing from case (i) to (ii) would be possible in an
interferometer having $h=2 w$. In that mode, the MZ interferometer
acts like a quantum negator for the channel that is transmitted
through the shorter interferometer arm. Interestingly, this
function is not simply performed here by a Rashba--induced spin
precession that was suggested earlier~\cite{egues:prl:02,pope:03}
as a means to induce phase shifts in a qubit. This is clearly
illustrated by the fact that the quantum negation in our MZ
interferometer occurs for electrons incident in any spin state, in
particular, in Rashba {\em eigenstates} that would not precess in
the previously discussed~\cite{pope:03} two--terminal device.

As a further application, the spin--dependent MZ interferometer
would be an excellent tool for measuring the effect of any
two--terminal device on carrier spin. Such a device could be
inserted into one of the interferometer arms, and changes in the
output conductances would directly reflect any spin flip or
rotation rendered by that device. This would be analogous to the
common use of optical MZ interferometers for measuring, e.g., the
refractive index of unknown materials.

The the above--discussed functions of the spin--dependent MZ
interferometer are independent of electron energy, rendering it
unnecessary to keep the electron density constant when tuning spin
splitting. Also, multi--subband devices can be expected to work
just as well as the single--subband case discussed here, as long
as the width of the quasi--onedimensional interferometer arms is
smaller than $L_{\text{so}}$. 
Realization of the suggested spin--dependent MZ interferometer
ultimately rests with the possibility to achieve electrostatic
control of quantum interference. Recent observation of the
electromagnetic Aharonov--Bohm effect~\cite{leokouwen},
measurement of voltage--controlled conductance modulation in an
electronic MZ interferometer~\cite{electronMZ} and, in particular,
demonstration of gate--controlled spin--orbit quantum interference
effects in lateral transport~\cite{cmarcus} suggest that the
associated experimental challenges can be met.

In conclusion, we have calculated transport properties of a
spin--dependent electronic Mach--Zehnder interferometer.
The interplay of electron--wave interference and Rashba spin
splitting results in a host of interesting electronic--transport
effects. In particular, this structure can work as a field--effect
switch. While the device operation is based on spin--dependent
interference effects, switching occurs independently of the spin
polarization of charge carriers and in the absence of magnetic
fields. Certain realizations of such an interferometer perform
quantum--logic gate functions. This example leads us to believe
that electronic analogs of other optical interferometers in
nanostructures with Rashba spin splitting will also lead to
interesting new spintronics devices with possible application in
quantum--information processing.



\end{document}